\documentclass[12pt]{article}
\usepackage{pic03}
\usepackage{cite}
\usepackage{subfigure}
\usepackage{graphicx}

\begin{document}

\title{\bf MEASUREMENT OF THE ATMOSPHERIC MUON SPECTRUM FROM 20~TO 2000 GeV}
\author{
Michael Unger \\
{\em DESY Zeuthen, Germany} \\
for the L3 Collaboration}
\maketitle

\baselineskip=14.5pt
\begin{abstract}
The atmospheric muon spectrum between 20 and 2000 GeV was measured with
the  L3 magnetic muon spectrometer 
for zenith angles ranging from 0 to 58 degrees. Due to the large data set
and the good detector resolution, a precision of 2.6\% at
100~GeV was achieved for the absolute normalization of the vertical 
muon flux. 
The momentum dependence of the ratio of positive to negative muons was
obtained between 20 and 630~GeV.
\end{abstract}

\baselineskip=17pt

\section{Introduction}
Atmospheric muons are amongst the final products of 
primary cosmic ray induced air shower cascades.
A precise measurement of the ground level muon flux
can therefore be used to test the understanding of the primary cosmic ray flux and
 the hadronic interactions involved in the production of the muons' parent mesons.
Moreover, it provides a crucial test for theoretical neutrino flux 
calculations, because each muon is produced with an accompanying muon neutrino.\\
Here a new measurement of the atmospheric muon flux is presented
using the precise muon spectrometer of the L3 detector~\cite{L3} located
at the LEP accelerator at CERN, Geneva. 
During the analysis special attention was given to the precise determination of all relevant
detector and environmental parameters needed to convert the raw data distributions to 
an absolute surface level flux. Due to the large amount of available statistics,
extensive studies of the residual systematic uncertainties were possible.
\section{Experimental Setup}
The momentum distribution of atmospheric muons is measured 
with the extended L3 setup known as L3+C~\cite{L3C}. 
It is located 450~m above sea level and shielded from the hadronic and electromagnetic air shower components
 by a 30~m thick molasse overburden.
The muon momentum is measured with the L3 muon spectrometer, which
is situated inside a large magnetic
volume of \mbox{1000 m$^3$} at a field of \mbox{0.5 T}. 
A 202  m$^2$ scintillator array was installed on top of the detector to record
the arrival time of the muons.
With this arrangement a relative momentum resolution ranging from 2.2\% at 20~GeV to
52\% at 2000~GeV is achieved.
Being equipped with a trigger and data acquisition system independent of 
the normal L3 data taking, L3+C recorded 1.2$\cdot10^{10}$ atmospheric muon
triggers during its operation in the years 1999 and 2000. 
\section{Analysis}
In total 2$\cdot10^7$ high quality events are used in the muon spectrum analysis. The geometrical acceptance
of the detector as well as the energy loss in the molasse overburden is calculated with a  simulation
of the L3+C setup and its surroundings. The detector efficiencies are determined from the data itself as a function of time, charge, momentum and zenith angle. 
The raw event
distributions at the detector level are deconvoluted taking into account the detector resolution and the stochastic
energy loss in the molasse. The systematic uncertainties of the muon flux and charge ratio measurement are studied by investigating the
stability of the results with time and azimuth angle and under the variation of the selection cuts.
\begin{figure}[t!]
\centering
       \subfigure[\itshape{Muon flux}]{\label{fig:spec}\includegraphics[width=0.48\linewidth] {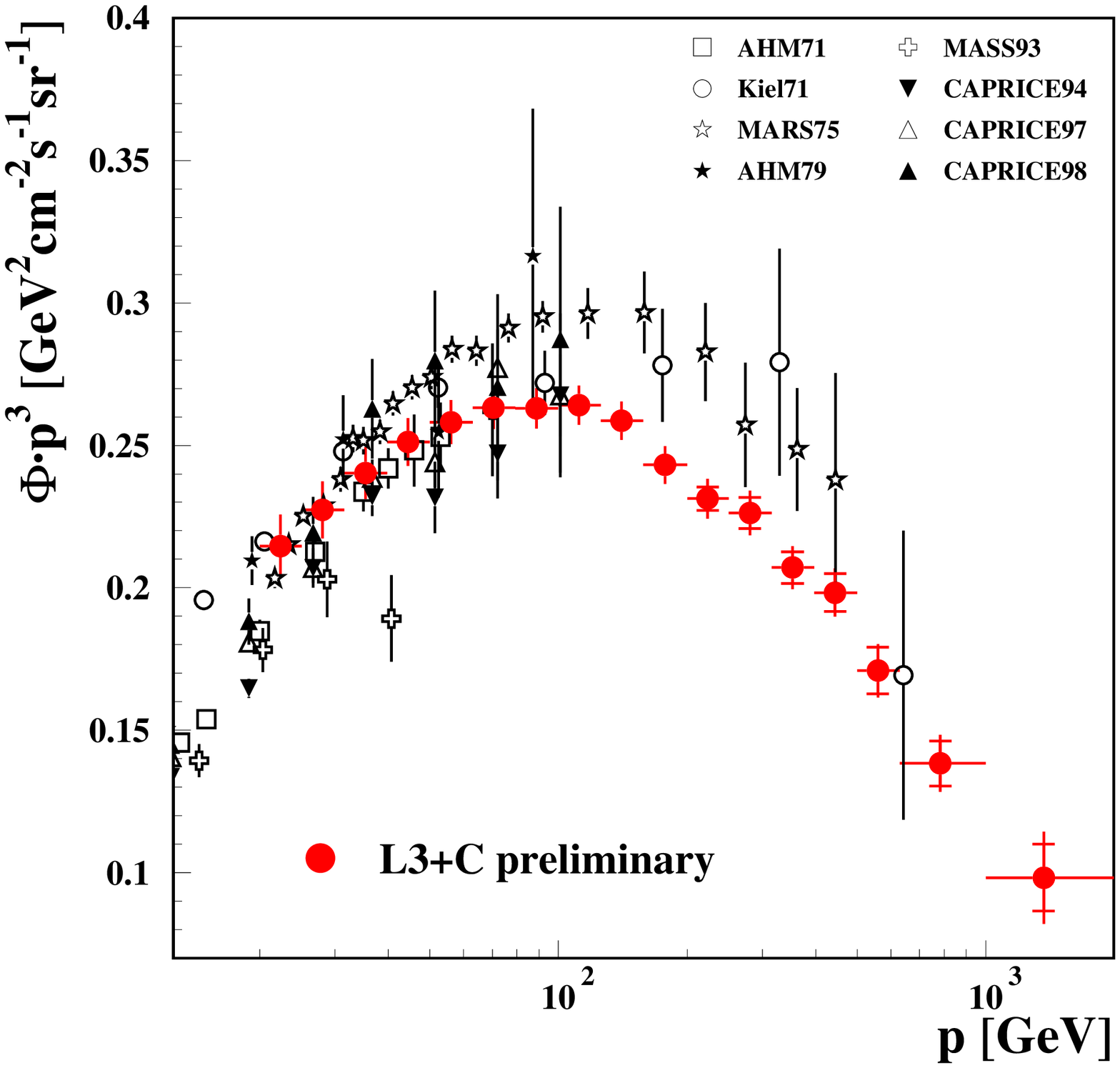}}
       \subfigure[\itshape{Charge ratio}]{\label{fig:crat}\includegraphics[width=0.48\linewidth] {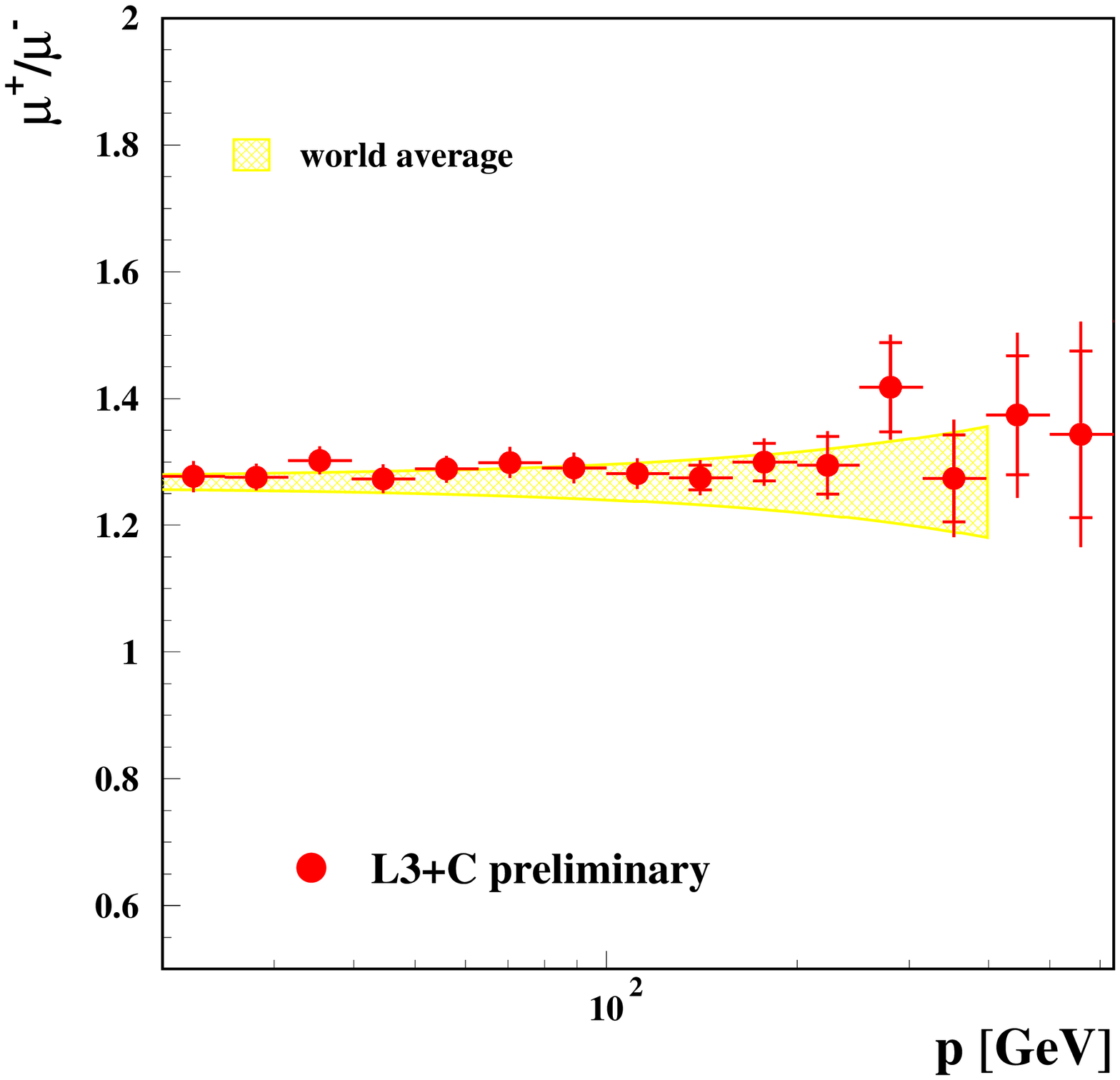}}
        \caption{\itshape{The L3+C vertical muon flux and charge ratio compared 
          to previous measurements~\cite{muexp,THCT}. 
The inner error bar denotes the statistical
error, the full error bar is the total error.}}
        \label{fig:result}
\end{figure}
\section{Results}
The L3+C vertical muon spectrum is shown in figure~\ref{fig:spec} along with previous measurements~\cite{muexp} providing an independent 
absolute normalization. The best precision is achieved around 100~GeV, where the total error amounts to~2.6\%. At lower
energies uncertainties of the molasse overburden get important, whereas at high energies the statistical error dominates.
The measured vertical charge ratio is compared to an average~\cite{THCT} of previous measurements in figure~\ref{fig:crat}. 
 It is worthwhile pointing out, that the total error of one zenith angle bin from this experiment
is compatible with the error from the average of all previous experiments.
Both, muon flux and charge ratio were measured as function of the zenith angle from 0 to 58 degrees. 
Their relative angular dependence is found in good agreement with the prediction from an air shower simulation~\cite{target} of
atmospheric muons.

\end{document}